УДК 517.5

Є.В. Сорока, студентка,
Б.В. Рубльов, д.ф.-м.н., проф.

Y.V. Soroka, student,
B.V. Rublyov, PhD, prof.

## Поріг вимирання популяції у просторовій стохастичній моделі

## Extinction threshold of a population in spatial and stochastic model

Київський національний університет імені Тараса Шевченка, Україна, 01601, місто Київ, вул. Володимирська, 64/13
e-mail: eugenia.soroka@gmail.com

Taras Shevchenko National University of Kyiv, 64/13, Volodymyrska Street, City of Kyiv, Ukraine, 01601
e-mail: eugenia.soroka@gmail.com

*У праці досліджено просторову стохастичну логістичну модель динаміки популяції певного виду. Досліджено критичні значення параметру смертності індивідуумів і знайдено поріг вимирання популяції за допомогою аналітичних методів і проведення симуляцій динаміки популяції. Порівняно аналітичні апроксимації порогу вимирання популяції з результатами симуляцій, на базі отриманих даних знайдені критичні значення параметрів смертності, при перевищенні яких відбувається зниження густоти популяції з подальшим її вимиранням. Досліджено поведінку критичного значення (порогу вимирання) як функції від параметрів моделі, зокрема, від просторового масштабу функцій конкуренції між індивідуумами та їхньої дисперсії. Перевірено гіпотезу про апроксимацію порогу вимирання аналітично і знайдено функціональну форму залежності критичного значення параметру смертності від параметрів моделі, для якої дійсна аналітична апроксимація.*

***Ключові слова:*** *просторова стохастична логістична модель, поріг вимирання, просторові кумулянти, збурюване розширення, модель середнього поля.*

*In this study, spatial stochastic and logistic model (SSLM) describing dynamics of a population of a certain species was analysed. The behaviour of the extinction threshold as a function of model parameters was studied. More specifically, we studied how the critical values for the model parameters that separate the cases of extinction and persistence depend on the spatial scales of the competition and dispersal kernels.*

*We compared the simulations and analytical results to examine if and how the mathematical approximations break down at the vicinity of the extinction threshold, and found a functional form of the naïve approximation for which higher-order term of analytical approximation converges.*

Статтю представив д.ф.-м.н., професор Хусаінов Д.Я.

### Introduction

One of the most fundamental questions in population biology concerns the persistence of species and populations, or conversely their risk of extinction. Theoretical ecologists have long sought to understand how the persistence of populations depends on biotic and abiotic factors. Extinction risk is influenced by a myriad of factors, including interaction between species traits and various stochastic processes leading to fluctuations and declines in population size [1].

The term extinction threshold refers to a critical value of some attribute, such as the amount of habitat in the landscape, below which a population, a metapopulation or species does not persist [2]. To calculate the extinction threshold, a model is required that links the relevant properties of the landscape to the dynamics of the species.

Mathematical and statistical modelling approaches provide a powerful tool for developing general theory and synthesizing the results of individual empirical case studies by placing them into the context of theory [3]. In particular, several modelling approaches have been developed to understand the ecological and evolutionary phenome related to extinction. In general, mathematical frameworks for dynamical systems can be classified depending on such parameters as type of a variable (discrete or continuous), space (accounted for in the model or not, or discrete or continuous), time (discrete or continuous) and stochasticity (accounted for or not).

In this study stochastic and spatial logistic model (SSLM) is used, and it is described as follows.

**Individual-based stochastic and spatial model**

Individual-based stochastic and spatial models form the most realistic family of population models, but at the same time they also are the most difficult family of models for mathematical analyses. In such models movements and interactions can be of localized nature, and demographic stochasticity is not averaged out.

Many individual-based models are defined with discrete spatial structure, e.g. a regular grid. Such models are also known as "interacting particle systems" or "stochastic cellular automata". Here we however focus on individual-based models formulated in continuous space and time. In particular, the spatial and stochastic logistic model is a spatio-temporal point process, the Lagrangian (individual-based) description of which is as follows:

Sedentary individuals produce propagules at a *per capita* fecundity rate *f* (by "rate" we mean probability per time unit, so that in a continuous-time model, the probability of a propagule being produced by a particular individual during a short time $dt$ is $fdt$). A newly produced propagule is distributed (instantaneously) according to a dispersal kernel, and it is assumed to establish (instantaneously) as a new-born individual, which matures (instantaneously) and starts to produce propagules. Existing individuals may die for two reasons. Firstly, there is a constant background per capita mortality rate $m$, yielding an exponentially distributed lifetime with mean $\frac{1}{m}$. Secondly, mortality has a density-dependent component (self-thinning), so that competition among the individuals may also lead to death. The density-dependent component of the death rate of a focal individual is a sum of contributions from all the other individuals within the entire $R^d$, but the strength of the competitive effect decreases with distance [4].

Next let us define SSLM mathematically. For this we consider the space of locally finite configurations:
$$\Gamma = \{\gamma \subset R^d \mid |\gamma \cap \Lambda| < \infty,$$
$$\text{for any bounded } \Lambda \subset R^d\} \quad (1.1)$$
where $d$ is the dimension of the space.

We use a probability measure $\mu_t$ (or $\mu(t)$) on $\Gamma$ to describe the state of the system at time $t$. Informally, the measure $\mu_t$ describes how likely the system is to be in a given configuration at time $t$, given that it starts from an initial state described by the measure $\mu_0$ at time 0. We define the SSLM by describing how individual events modify an observable $F$. More precisely, the evolution of states is defined through the differential equation
$$\frac{d}{dt}\langle F, \mu(t)\rangle = \langle LF, \mu(t)\rangle, \quad (1.2)$$
where $L$ is a linear operator acting on observables, i.e., functions on $\Gamma$ [4]. Here $\langle F, \mu(t)\rangle$ is a pairing between an observable and a measure, defined by
$$\langle F, \mu(t)\rangle := \int_\Gamma F(\gamma) d\mu(\gamma). \quad (1.3)$$

As for the evolution of states associated with SSLM, this model can be mathematically defined through the linear operator $L$ with
$$(LF)(\gamma) =$$
$$\sum_{x \in \gamma}\left(m + \sum_{y \in \eta \setminus x} a^-(x-y)\right)[F(\gamma \setminus x) - F(\gamma)] +$$
$$+ \sum_{y \in \gamma}\int_{R^d} a^+(x-y)[F(\gamma \cup x) - F(\gamma)]\,dx \quad (1.4)$$

where $F$ is arbitrary observable; $m$ is the density-independent death rate of individual in location $x$; $a^-(x-y)$ is a kernel describing the mortality rate imposed by an individual located at $y$ to individual located in $x$; the reproduction kernel $a^+(x-y)$ indicates the rate (per unit area) at which new-born individuals are created at location $x$ by a parent located at $y$. In this model per capita fecundity rate $f$ is incorporated in the reproduction kernel $a^+$, i.e. $f = \int_{R^d} a^+(x)dx$ [4].

**Aim of the work**

The aim of this work is to analyse the behaviour of the extinction threshold in the spatial stochastic and logistic model. Recent mathematical developments enable one to study the dynamics of this model using a mathematically rigorous approximation, namely a perturbation expansion around the mean-field model (for details, see below). However, the perturbation expansion may break down at the vicinity of the extinction threshold, and thus there is yet no mathematical theory that could be used to understand the dynamics of population extinction and the behaviour of the extinction threshold as a function of model parameters. The aim of this work is to examine these questions with the help of simulations.

More specifically, we aim to study how the critical values for the model parameters that separate the cases of extinction and persistence depend on the spatial scales of the competition and dispersal kernels. A further aim is to compare the simulations to analytical results to examine if and how the mathematical approximations break down at the vicinity of the extinction threshold. We formulate the results as a mathematically formulated conjecture, which we hope will inspire future mathematical work, that may eventually lead to its confirmation or rejection.

We start by reviewing some relevant theoretical results that have been published earlier. We then present the methods used in this work, then the results obtained, and finally the conclusions of this work.

**Review of previously published theory**
*Spatial moments and cumulants*

While the operator $L$ (together with initial measure $\mu_0$ – the probability distribution for the initial distribution of individuals in the domain) defines the model, as such it yields no statistical information on how the population behaves. To analyse the model behaviour, we turn to the time-evolution of spatial moments and spatial cumulants, which translate the Lagrangian (individual-based) model definition into an Eulerian (population-based) framework.

The $n^{th}$ order spatial moment is denoted by the function $k^{(n)}(x_1, x_2, \ldots, x_n)$. It is assumed to be symmetric, and the spatial moments of all orders $n = 0, 1, \ldots$ are collected into the family $k = \{k^{(n)}\}$, with $k(0) = 1$. The vector $k$ of all spatial moments is a sufficient description of the state of the system, i.e., it includes the same statistical information as the

The first order spatial moment describes expected population density, while the second and higher orders describe the degree of clustering in the spatial distribution of individuals [4]. In the case of Poisson measure (i.e., complete spatial randomness), the spatial moment function of any order $n$ is simply given by the product

$$k^{(n)} = (x_1, x_2, \ldots, x_n) = \prod_{i=1}^{n} k^{(1)}(x_i). \quad (2.1)$$

Spatial cumulants can be defined as [4]
$$\begin{aligned} u^{(0)} &= 0, \\ u^{(1)}(x) &= k^{(1)}(x), \quad (2.2) \\ u^{(2)}(x_1, x_2) &= k^{(2)}(x_1, x_2) - u^{(1)}(x_1)u^{(1)}(x_2) \end{aligned}$$
and, for any $\eta$, $|\eta| = n \geq 2$,
$$u^{(n)}(\eta) := k^{(n)}(\eta) - \sum_{\substack{\eta_1 \sqcup \ldots \sqcup \eta_s = \eta \\ s \geq 2, \eta_i \neq \emptyset \\ for\ i=1,\ldots,s}} \frac{1}{s!} u(\eta_1) \ldots u(\eta_s).$$

Thus, as in the non-spatial case, the cumulant of order $n$ is obtained from the moment of order $n$ by subtracting all combinations of lower order cumulants.

The dynamical equation for the first spatial cumulant (population density) for SSLM reads

$$\begin{aligned} \frac{d}{dt} u_t^{(1)}(x) = &-m u_t^{(1)}(x) \\ &- \int_{\mathbb{R}^d} a^-(x-y) u_t^{(2)}(y, x)\, dy \\ &+ \int_{\mathbb{R}^d} a^+(x-y) u_t^{(1)}(y)\, dy \\ &- u_t^{(1)}(y) \int_{\mathbb{R}^d} a^-(x-y) u_t^{(1)}(y)\, dy, \quad (2.4) \end{aligned}$$

where the last term in the right-hand side is the non-linear component.

As for the spatial moment equations, spatial cumulants form an infinite hierarchy that can not be solved exactly. However, unlike spatial moments, spatial cumulants of higher orders can be expected to be small in the sense that they tend to zero as the distance between any two points in the definition tends to infinity.

Next we discuss a perturbative approach that utilizes this property of the cumulants.

*Analytical approximations of spatial cumulants*

For approximation of first and second order spatial cumulants at the equilibrium, we use the scaled version of the model, where the scaling parameter $\varepsilon$ is an arbitrary positive number ($\varepsilon > 0$). As in the case for the SSLM, operator $L$ may include functions (kernels) which describe pair interactions between the elements of the system. For example, in case of the SSLM, two such kernels are involved, namely $a^-$ (competition / density-dependent mortality) and $a^+$ (reproduction and dispersal). For any such kernel $a: \mathbb{R}^d \to \mathbb{R}_+ := [0, +\infty)$, $a \in L^1(\mathbb{R}^d)$, we define a scaled version $a_\varepsilon$ by setting

$$a_\varepsilon(x) := \varepsilon^d a(\varepsilon x), \quad x \in \mathbb{R}^d. \quad (2.5)$$

Then, as $\varepsilon \to 0$, the kernel becomes increasingly flat and long-ranged, while its integral remains constant, i.e.

$$\int_{\mathbb{R}^d} a_\varepsilon(x)\, dx = \int_{\mathbb{R}^d} a(x)\, dx \quad (2.6)$$

independently of $\varepsilon > 0$. For a given model defined by an operator $L$, we define a scaled model by replacing the operator $L$ by $L_\varepsilon$, meaning that all the kernels of $L$ are rescaled according to Eq. 2.6.

As $\varepsilon \to 0$ a so-called *mean-field (or mesoscopic) limit* can be obtained, which generally refers to a situation in which the law of mass action holds, i.e. it assumes that individuals are (at least locally) well-mixed in the sense that the probability of interaction of a randomly chosen individual with any other individual from the same population does not depend on the individual chosen [4]. The limit $\varepsilon \to 0$ corresponds to one such particular limit, which we call that of long-ranged interactions. In case of the SSLM model, one may develop a perturbation expansion for the dynamical equations of the first and second order spatial cumulants as

$$u_\varepsilon^{(1)}(t,x) = q(t,x) + \varepsilon^d p(t,x) + O(\varepsilon^{2d}), \quad (2.7)$$
$$u_\varepsilon^{(2)}(t,x,y) = \varepsilon^d g(t,x,y) + O(\varepsilon^{2d}). \quad (2.8)$$

The dynamics of $q$, $p$ and $g$ can be explicitly written down as a function of the model parameters [4] as follows.

Population density $q$ satisfies the differential equation

$$\frac{d}{dt} q(t,x) = -mq(t,x)$$
$$-q(t,x)\int_{\mathbb{R}^d} a^-(x-y)q(t,y)\,dy + \quad (2.9)$$
$$+\int_{\mathbb{R}^d} a^+(x-y)q(t,y)\,dy.$$

The usual non-spatial logistic model is obtained by further assuming *translational invariance*, i.e. that the initial condition is independent of spatial location. In this case, the functions $q$ and $p$ become independent of the location $x$, and the function $g$ depends only on the distance between $x$ and $y$ (denoted by $r = |x-y|$). Then the mean-field population density evolves as

$$\frac{d}{dt} q(t) = (A^+ - m)q(t) - A^- q(t)^2, \quad (2.10)$$

where $A^+ = \int_{\mathbb{R}^d} a^+(x)dx$ and $A^- = \int_{\mathbb{R}^d} a^-(x)dx$ denote the integrals of the reproduction and mortality kernels, respectively.

For functions $p(t)$ and $g(t,r)$ it holds:

$$\frac{dg}{dt} = 2\, g * a^+ + 2qa^+ - 2mg - 2\, g * a^-$$
$$- 2\, qA^- g - 2q^2 a^-, \quad (2.11)$$
$$\frac{dp(t)}{dt} = A^+ p(t) - mp(t) - 2A^- q(t)p(t)$$
$$- \int_{\mathbb{R}^d} a^-(x)g(x)dx, \quad (2.12)$$

where $*$ denotes convolution. The convergence of the SSLM to the mean-field has been rigorously proved earlier in [5]. [4] shows that the first order correction term (the coefficient of $\varepsilon^d$) is non-zero only on the space of one- and two-point configurations (denoted by $p(t)$ and $g(t,r)$, respectively). Thus, for large but finite interactions, the two-point spatial cumulant dominates the spatial pattern, the higher order cumulants being less important.

As time parameter $t$ tends to infinity, the dynamics of the system and hence also $u^{(1)}(t)$ and $u^{(2)}(t,r)$ can be expected tend to a stationary state, which we denote by

$$u^{(1)*} = \lim_{t \to \infty} u^{(1)}(t), \quad (2.13)$$
$$u^{(2)*}(r) = \lim_{t \to \infty} u^{(2)}(t,r). \quad (2.14)$$

Let us define by $q^*$, $g^*$ and $p^*$ the corresponding values of $q(t)$, $g(t,r)$ and $p(t)$ at the equilibrium to which the dynamics can be expected to converge as $t \to \infty$. These can be solved explicitly from the dynamical equations given above, i.e.

$$q^* = \frac{A^+ - m}{A^-}, \quad (2.15)$$
$$\tilde{g}^* = \frac{q^{*2}\tilde{a}^- - q^*\tilde{a}^+}{\tilde{a}^+ - \tilde{a}^- - q^*A^- - m}, \quad (2.16)$$
$$p^* = \frac{\int_{\mathbb{R}^d} a^-(x)g(x)dx}{A^+ - m - 2A^- q^*}$$
$$= -\frac{\int_{\mathbb{R}^d} a^-(x)g(x)dx}{A^+ - m}, \quad (2.17)$$

where $\tilde{f}$ denotes Fourier transform of function $f$.

**Method Section**
*Analytical methods*

In this study the behaviour of the critical value $m_c(\varepsilon)$ for the persistence of population as a function of $\varepsilon$ is being investigated. If we simply ignore the error term in Eq. 2.7, $m_c(\varepsilon)$ can be obtained by solving the equation

$$q^* + \varepsilon^d p^* = 0, \quad (3.1)$$

where $q^*$ and $p^*$ are the values of $q(t)$ and $p(t)$ at the equilibrium, which depend on mortality rate $m$ (by definition). Let us call such $m_c(\varepsilon)$ a "naïve approximation", and it can be written down explicitly as

$$m_c(\varepsilon) = A^+ - \varepsilon^{\frac{d}{2}}\sqrt{A^- \int_{\mathbb{R}^d} a^-(x)g(x)dx}. \quad (3.2)$$

The naïve approximation (given by Eq. 3.2) defines a critical value $m_c(\varepsilon)$, such that the persistence criteria is $m_\varepsilon < m_c(\varepsilon)$, i.e. as $m_\varepsilon \geq m_c(\varepsilon)$ population goes extinct.

The reason why the naïve approximation may not hold is the following. Let us write the perturbation expansion (see Eq. 2.7) as

$$u^{(1)*} = q^* + \varepsilon^d p^* + C(m)\varepsilon^{2d} + O(\varepsilon^{3d}). \quad (3.3)$$

While the constant $C(m)$ is finite for any fixed $m$, it may diverge as $C(m) \to \infty$ when the parameter $m$ approaches the extinction threshold determined by the mean-field model, i.e. when $m \to m_c(0) = A^+$. If and how this happens can make the naïve approximation invalid. Another reason why the naïve approximation may not hold is the presence of the higher order terms included in $O(\varepsilon^{3d})$, as their coefficients may also diverge as $m \to m_c(0)$.

To simplify, we ignore here the higher order terms, and thus assume

$$u^{(1)*} = q^* + \varepsilon^d p^* + C(m)\varepsilon^{2d}. \quad (3.4)$$

To explore how fast $C(m)$ can diverge so that the naïve approximation is still valid, we assume that
$$C(m) = O(A^+ - m)^{-x}, \quad (3.5)$$
where $x > 0$ so that the term diverges as $m \to A^+$. If the naïve approximation holds, at the vicinity of the extinction threshold it holds that $A^+ - m = O(\varepsilon^{d/2})$, and thus the term $\varepsilon^d p^*$ behaves as $O(\varepsilon^{d/2})$, whereas the term $C(m)\varepsilon^{2d}$ behaves as $O\left(\varepsilon^{2d-x\frac{d}{2}}\right)$. The naïve approximation is expected to hold if the latter term decreases faster than the former term, and thus if $2d - x\frac{d}{2} > \frac{d}{2}$. This equation holds if $x < 3$. In other words, if $x < 3$, the naïve approximation is expected to describe the leading behaviour of $m_c(\varepsilon)$ for small $\varepsilon$. We however note that this reasoning ignores the higher order terms in the perturbation expansion.

The above considerations motivate us to conduct simulations to find out about the behaviour of $m_c(\varepsilon)$ for small $\varepsilon$, and the behaviour of $C(m)$ near the extinction threshold.

But first let us define three errors useful for studying and understanding the differences between analytical and simulated results, and also for obtaining formula for $m_c(\varepsilon)$:
$$e_1 = -p^*\varepsilon^d, \quad (3.8)$$
$$e_2 = -\left(u_\varepsilon^{(1)*} - q^*\right), \quad (3.9)$$
$$e_3 = -(u_\varepsilon^{(1)*} - (q^* + p^*\varepsilon^d)). \quad (3.10)$$

According to the theory reviewed above, error $e_2$ should behave for small epsilon as $e_1$ and thus as $O(\varepsilon^d)$, whereas error $e_3$ should behave as $O(\varepsilon^{2d})$ because $u^{(1)*} = q^* + \varepsilon^d p^* + O(\varepsilon^{2d})$ (see Eq. 2.7 and Eq. 3.3).

*Numerical methods*

As noted above, the perturbation expansion may break down close to the extinction threshold, and thus there is yet no mathematical theory that could be used to understand the dynamics of population extinction and the behaviour of the extinction threshold as a function of model parameters.

In order to approach this problem, we approximate first and second order spatial cumulants at the equilibrium by simulating the dynamics of the population defined by SSLM.

We define the model with the following parameters:

1. Density independent birth kernel $a^+(x)$ is as a tophat kernel with parameters integral and radius $(A^+, \frac{1}{\varepsilon})$, where $A^+ = 2$.

2. Death by competition kernel $a^-(x)$ is also defined as a tophat kernel with parameters $(A^-, \frac{1}{\varepsilon})$, where $A^- = 1$.

3. Mortality (density independent death) rate $m$ is a constant which takes values from the set $\{0.5, 0.6, 0.7, \ldots, 1.7, 1.8\}$.

We conduct simulations for the perturbation parameter $\varepsilon$ covering the range $\varepsilon \in \{1, \frac{1}{2}, \frac{1}{4}, \frac{1}{8}, \frac{1}{16}, \frac{1}{32}, \frac{1}{64}\}$.

Population dynamics are simulated on a torus domain of size $U \times U$ during time $T$, where as a starting point $(U, T) = (128, 128)$. We then sequentially increase either the final time of the simulation $T$ or the domain size $U \times U$ by fourfold, continuing until the inferred qualitative behaviour of the extinction threshold does not change anymore, and also its quantitative behaviour remains essentially unchanged.

For approximating $u^{(1)*}$ and $u^{(2)*}(r)$ by simulations, we define $u_{U,T}^{(1)}$ which denotes the mean density obtained by simulating the dynamics in a domain of size $U \times U$ until final time $T$, and $u_{U,T}^{(2)}(r)$ analogously, i.e. it denotes the second order cumulant for a population in $U \times U$ domain until final time $T$.

We also define $u_{U,T}^{(1)*}$ as a limit of mean density obtained by simulations in a domain of size $U \times U$ until time $T$, calculated discarding the first half $T/2$ of time of the simulation and averaging the recorded density over the rest of time. Analogously, $u_{U,T}^{(2)*}(r)$ denotes the second order cumulant for a population in $U \times U$ domain until final time $T$, discarding the first half of time, and averaging the values over the rest of time.

However, we note that approximating $u^{(1)*}$ and $u^{(2)*}(r)$ by simulating the system is not trivial. This is because for a fixed domain size $U$ the population always goes extinct for $T \to \infty$. Conversely, for a finite final time $T$ some individuals are always predicted to remain in the population for $U \to \infty$.

*Predicting the extinction threshold*

To develop our approach for studying the extinction threshold from simulations, let us return to the naïve approximation. In this case the critical value of $m_c(\varepsilon)$ can be solved from Eq. 3.1, or
$$u_\varepsilon^{(1)A}(m) := q^*(m) + \varepsilon^d p^*(m) = 0, \quad (3.11)$$
where the behaviour of both $q^*$ and $p^*$ is analytically known (see above), and the superscript A in $u_\varepsilon^{(1)A}$ refers to the fact that $u_\varepsilon^{(1)A}$ is the analytical approximation of $u_\varepsilon^{(1)}$.

Now, let us pretend that we would not know the behaviour of $p^*$ as a function of $m$, only that of the mean-field model $q^*(m)$, but that we would know the values of $u_\varepsilon^{(1)A*}(m)$ for a collection of values of $m$ and $\varepsilon$. As we defined the error $e_1$ as $-p^*\varepsilon^d$ (Eq. 3.8), it holds that $e_1 = q^*(m) - u_\varepsilon^{(1)A*}(m)$. Considering these as data, we may fit a statistical model, and thus infer the behaviour of $e_1$ as a function of $m$ and $\varepsilon$, and consequently the behaviour of the extinction threshold $m_c(\varepsilon)$. We first develop this approach so that it provides an accurate approximation of the naïve approximation. After that, we apply the same idea for data based on the simulations.

Using Wolfram Mathematica, non-linear model is fitted to both errors $e_1$ and $e_2$. Knowing functional forms for errors $e_1$ and $e_2$, we can obtain critical value for mortality $m_c(\varepsilon)$ by solving the equation for population extinction at the equilibrium:

$$u_\varepsilon^{(1)*} \approx q^* + \varepsilon^d p^* = 0, \qquad (3.12)$$

and thus predict the extinction threshold.

For error $e_1 = -p^*\varepsilon^d$ functional form $e_1 = f(m) \cdot \varepsilon^d$ is fitted, where $f(m) = \frac{a + b \cdot (A^+ - m)}{1 + c \cdot (A^+ - m)}$, and values of parameters $a, b$ and $c$ are estimated. In order to conduct such model fitting, data for error $e_1$ is log-transformed, and on the right-hand side of the equation Taylor series expansion of $\log_2(f(m) \cdot \varepsilon^d)$ is used.

After model fitting we obtain functional form of error $e_1$, and therefore can calculate mortality $m_c(\varepsilon)$ by solving the equation:

$$u_\varepsilon^{(1)*} \approx q^* - e_1 = 0; \qquad (3.13)$$

$$u_\varepsilon^{(1)*} \approx \frac{A^+ - m}{A^-} - f(m) \cdot \varepsilon^d = 0. \qquad (3.14)$$

For error $e_2 = -(u_\varepsilon^{(1)} - q^*)$ (Eq. 3.9) first model $e_2 = f(m) \cdot \varepsilon^d$ is fitted, where $f(m) = \frac{a + b \cdot (A^+ - m)}{1 + c \cdot (A^+ - m)}$, and values of parameters $a, b$ and $c$ are estimated. Then model $e_2 = f(m) \cdot \varepsilon^d + g(m) \cdot \varepsilon^{2d}$ is fitted, where $g(m) = \frac{a' + b' \cdot (A^+ - m)}{1 + c' \cdot (A^+ - m)}$, and values of $a, b, c, a', b'$ and $c'$ are estimated. In the second case model fits better, which displays that higher-order term $\varepsilon^{2d}$ is important for approximating $e_2$. The same procedure as with error $e_1$ is used. After model fitting with the second functional form, we can calculate $m_c(\varepsilon)$ by solving the equation:

$$u_\varepsilon^{(1)*} \approx q^* - e_2 = 0; \qquad (3.15)$$

$$\frac{A^+ - m}{A^-} - \big(f(m) \cdot \varepsilon^d + g(m) \cdot \varepsilon^{2d}\big) = 0. \qquad (3.16)$$

For error $e_3 = -(u_\varepsilon^{(1)} - (q^* + \varepsilon^d p^*))$ (Eq. 3.10) model $e_3 = f(m) \cdot \varepsilon^{2d}$ is fitted, where $f(m) = \frac{a + b \cdot (A^+ - m)}{1 + c \cdot (A^+ - m)}$, and values of parameters $a, b$ and $c$ are estimated. Then we calculate $m_c(\varepsilon)$ by solving the equation:

$$u_\varepsilon^{(1)*} \approx q^* + \varepsilon^d p^* - e_3 = 0; \qquad (3.17)$$

$$u_\varepsilon^{(1)*} \approx \frac{A^+ - m}{A^-} - f(m) \cdot \varepsilon^{2d} = 0. \qquad (3.18)$$

For errors $e_2$ and $e_3$ the mean values of the data from the repeated simulations are used for model fitting, incorporating weights $\frac{1}{error^2}$ into non-linear model, where $error$ is a deviation for each measurement.

**Results**

To illustrate how results obtained by simulations correspond to those obtained analytically, Fig. 1 shows the comparison of simulated and analytical population dynamics for the first order (a) and second order (b) spatial cumulants.

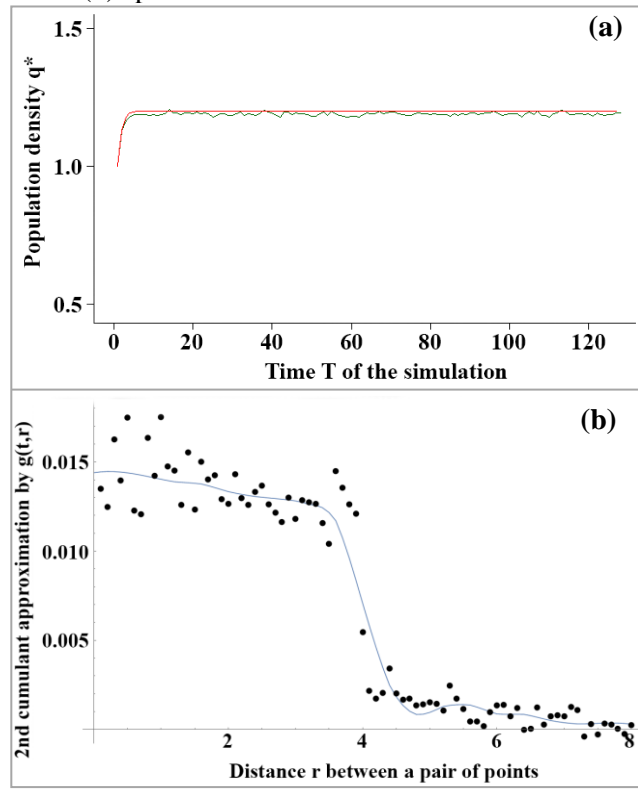

Fig. 1 – Population density (a) and $2^{nd}$ spatial cumulant (b) for the simulation with domain size $U = 256$, time of the simulation $T = 128$, mortality rate $m = 0.8$ and scaling parameter $\varepsilon = \frac{1}{4}$. In (a) analytical solution is shown in red and results of the simulation in green. In (b) analytical solution is the blue curve and results of the simulation are denoted by black dots.

Simulations enabled us to obtain necessary data on the dynamics of a population and conduct further analysis required for predicting the extinction threshold.

After fitting a functional model to the data from errors $e_1, e_2$ and $e_3$ we obtained estimated critical values of mortality rate $m_c(\varepsilon)$ (Fig. 2).

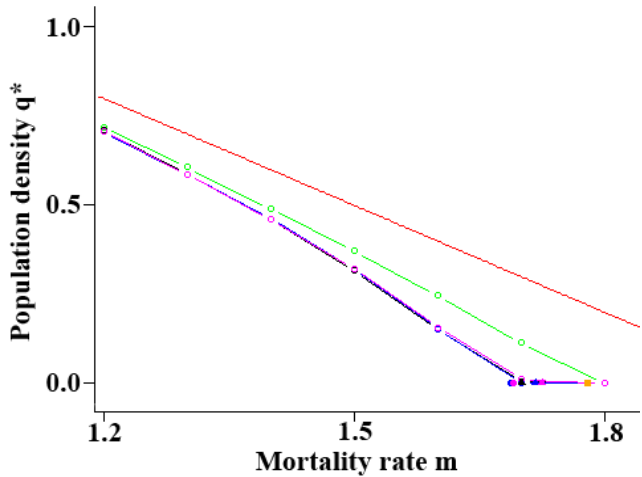

Fig. 2 – Population density: analytical in the mean-field model (red), numerical as $\varepsilon = \frac{1}{2}$ (green) and simulated: $(U, T) = (128, 128)$ (black), $(U, T) = (128, 512)$ (blue), $(U, T) = (256, 128)$ (magenta). Predicted values $m_c(\varepsilon)$ are listed as follows:

| (U, T) | MF to error | $m_c(\varepsilon)$ | Denoted by |
|---|---|---|---|
|  | $e_1$ | 1.7804 | 🟧 |
| (128, 128) |  | 1.6890 | ⚫ |
| (128, 512) | $e_2$ | 1.6876 | 🔵 |
| (256, 128) |  | 1.6914 | 🟣 |
| (128, 128) |  | 1.7011 | ▲ |
| (128, 512) | $e_3$ | 1.7174 | 🔺 |
| (256, 128) |  | 1.7253 | 🔺 |

(MF stands for model fitting.)

## Conclusions

In this study spatial stochastic and logistic model (SSLM) describing dynamics of a population of a certain species was analysed. Specific cases of SSLM with various model parameters were studied and the corresponding population dynamics were simulated and obtained data was analysed.

The behaviour of the extinction threshold as a function of model parameters was studied. More specifically, we studied how the critical values for the model parameters that separate the cases of extinction and persistence depend on the spatial scales of the competition and dispersal kernels defined in the model.

We compared the simulations and analytical results to examine if and how the mathematical approximations break down at the vicinity of the extinction threshold, and found a functional form of the naïve approximation for which higher-order term of analytical approximation converges, i.e. the functional form for which the mathematical approximation of spatial cumulants holds.

Further work should enable us to formulate the results as a mathematically formulated conjecture about the dependence of the extinction threshold on model parameters. We hope that such conjecture will awaken interest in the minds of mathematicians which would eventually lead to its rigorous mathematical confirmation or rejection.